# STONE-WALES DEFECTS IN NITROGEN-DOPED $C_{20}$ FULLERENES: INSIGHT FROM *AB INITIO* CALCULATIONS

Konstantin P. Katin[1,2] and Mikhail M. Maslov[1,2]

[1]Department of Condensed Matter Physics, National Research Nuclear University "MEPhI", Kashirskoe Shosse 31, Moscow 115409, Russia

[2]Laboratory of Computational Design of Nanostructures, Nanodevices and Nanotechnologies, Research Institute for the Development of Scientific and Educational Potential of Youth, Aviatorov Street 14/55, Moscow 119620, Russia

ABSTRACT

Density functional theory is applied to study the mechanism of the Stone-Wales defect formation in pure and nitrogen-doped dodecahedral $C_{20}$ fullerenes. The molecular structures of initial and defected cages as well as transition states dividing them are obtained. Depending on the number of nitrogen atoms and their relative position in the cage, Stone-Wales defect is formed through the single additional intermediate state or directly. The activation energy barrier of the defect formation reduces from 4.93 eV in pure $C_{20}$ to 2.98 eV in single-doped $C_{19}N$, and reaches ~2 eV under further doping. All nitrogen-doped fullerenes considered possess high kinetic stability at room temperature. However, they become much less stable at temperatures of about 750 K that are typical for the fullerene annealing process.

KEYWORDS: kinetic stability; transition state; DFT; $C_{19}N$; $C_{18}N_2$; $C_{17}N_3$.

Corresponding author: Konstantin P. Katin, Associate Professor at Department of Condensed Matter Physics, National Research Nuclear University "MEPhI", Kashirskoe Shosse 31, 115409 Moscow, Russian Federation, e-mail: KPKatin@yandex.ru.

INTRODUCTION

Nitrogen doping of pure carbon fullerenes provides the possibility for modification of their electronic properties and reactivity [1,2]. In addition, typical C–C distances in fullerenes are close to the C–N bond length in trimethylamine, which is equal to 1.451 Å [3]. Thus, the nitrogen atoms can be embedded into the fullerene cage instead of the carbons without perceptible structural deformations. Introduced nitrogens provide *n*-doping of fullerenes and significantly change their electronic characteristics. So, the nitrogen-doped fullerenes are the promising candidates for the active parts of nanoelectronic devices [1], as the cathode catalysts for hydrogen fuel cells [4], or as the adsorbents for Tabun nerve agent [5].

First nitrogen substituted derivatives of $C_{60}$ and $C_{70}$ were synthesized in 1991 via the contact-arc vaporization of graphite in a partial atmosphere of gaseous $N_2$ and $NH_3$ [6]. Next, a number of N-doped [60]fullerenes, including $C_{59}N$ [7], $C_{57}N_3$ [8], and $C_{48}N_{12}$ [9], were isolated as well. Successful synthesis of N-substituted fullerenes stimulated their further theoretical studies: energy, geometry and electronic properties were calculated for the N-substituted fullerenes $C_{19}N$ [10], $C_{20-n}N_n$ (*n* = 1 – 12) [11], $C_{34}N_2$ [12], and $C_{60-n}N_n$ (*n* = 1 – 12) [13]. Significant charge transfer between the carbon and nitrogen atoms was detected for both $C_{20-n}N_n$ (*n* = 1 – 12) [11] and $C_{60-n}N_n$ (*n* = 1 – 12) [13] molecular systems. Note that the all N-doped fullerenes considered were found to be the true minima on the corresponding potential energy surfaces (PES). Nevertheless, introduction of nitrogen atoms into the high-strained carbon cage can induce the defects formation at non-zero temperatures. For example, in (6,6) carbon nanotube nitrogen doping reduces the activation barrier for Stone-Wales defect [14] formation by 2.28 eV [15]. Thus, one can expect the similar or even greater reduction of the isomerization barriers for more curved N-doped fullerenes.

The smallest carbon fullerene $C_{20}$ is less kinetically stable than the higher fullerenes [16,17], due to its stronger curvature. The defect formation processes in $C_{20}$ were previously studied in the frame of semiempirical molecular dynamics simulations [18,19] and within the *ab initio* calculations of corresponding potential energy surfaces [16]. It was found that $C_{20}$ fullerene possesses surprisingly high kinetic stability [18]. Nevertheless, introducing the nitrogen atoms into the fullerene cage can trigger the defects formation in the cluster. Despite of the intensive studies of $C_{20-n}N_n$ systems [10,11] and promising perspectives of their application [4], kinetic stability and the processes of defect formation in these metastable compounds was not yet investigated. At the same time, structural defects can significantly change their energy and electronic properties as well as stability, as it was earlier shown for the Stone-Wales defects in fullerene $C_{36}$ [20], non-classical fullerene $C_{46}$ [21], nitrogen-doped carbon nanotubes [22], and graphene nanoribbons [23].

In this Article, we present a theoretical study of defects formation in pure and nitrogen-doped $C_{20-n}N_n$ ($n = 0 - 3$) fullerenes. The influence of one, two and three substitutional nitrogen atoms on activation energy barrier that should be overcome for dodecahedral cage transferring into its defective isomer is investigated in detail.

COMPUTATIONAL DETAILS

Our calculations are performed in the frame of density functional theory using the hybrid RO-B3LYP exchange-corrected functional [24,25] with the 6-311G(d,p) electronic basis set [26]. For $C_{20}$ isomerization we obtain the minimum energy path (MEP) using the nudged elastic band (NEB) method [27] as it is implemented in the TeraChem program package [28–31]. We find the local minima and transition states within the Newton-Raphson method. Optimization procedures are continued until the residual forces acting on atoms become lower than $10^{-4}$ Ha/bohr. For all located stationary points, the hessian matrix at the same level of theory is calculated and the frequency analysis is performed for confirming the true minima (there are no imaginary frequencies in the spectrum) or transition states (frequency spectrum contains only one imaginary frequency). Next, the intrinsic reaction coordinate analysis is performed to confirm that the located transition states separate the corresponding minima. All calculations excluding the nudged elastic band application are carried out using the GAMESS software [32].

RESULTS AND DISCUSSION

First of the all, we optimize the geometries of pure $C_{20}$ fullerene and its nitrogen-doped derivatives (see Figure 1).

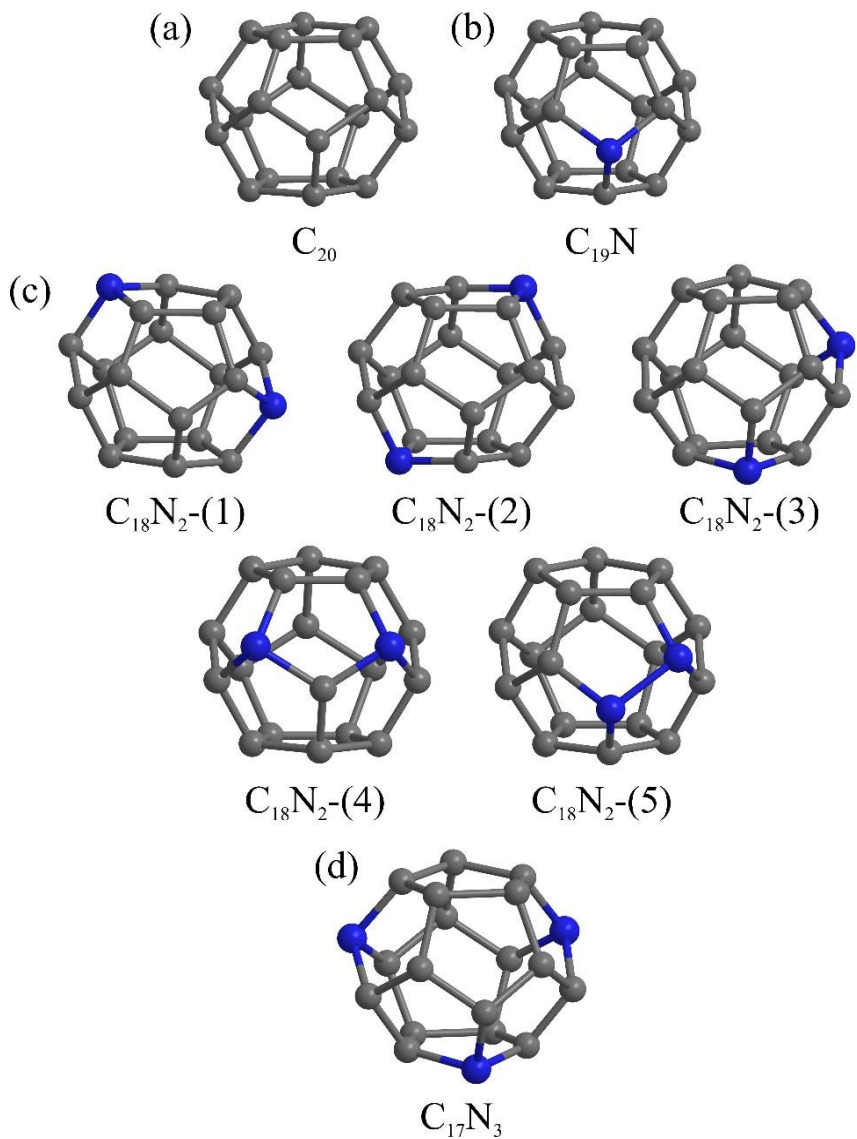

**Figure 1**. Molecular structures of (a) $C_{20}$, (b) $C_{19}N$, (c) $C_{18}N_2$, and (d) $C_{17}N_3$ cages. Grey and blue colors denote the carbon and nitrogen atoms, respectively.

Note that there are five various dodecahedral $C_{18}N_2$ isomers. Their molecular structures are presented at Figure 1c. It is found that the isomers $C_{19}N_2$-(4) and $C_{19}N_2$-(5) that have the closest relative position of nitrogen atoms are the most energetically unfavorable. Thus, we further consider the most energetically favorable $C_{18}N_2$ isomers labeled as $C_{18}N_2$-(1), $C_{18}N_2$-(2) and $C_{18}N_2$-(3) at Figure 1c and one of the $C_{17}N_3$ isomers, in which nitrogen atoms are maximally separated from each other (Figure 1d). Frequency analysis confirms that all obtained structures are the true minima on PES. Thermochemical data calculated for the normal conditions and for the temperature $T = 750$ K that is typical for annealing of nitrogen-doped fullerenes [8] are listed in the Table 1.

Table 1. Internal energies $E$ (kJ/mol), enthalpies $H$ (kJ/mol), and Gibbs energies $G$ (kJ/mol) of the $C_{20}$ fullerene and its nitrogen-doped derivatives. Pressure is equal to standard value (101.3 kPa).

|  | $T = 298$ K | | | $T = 750$ K | | |
|---|---|---|---|---|---|---|
| System | $E$ | $H$ | $G$ | $E$ | $H$ | $G$ |
| $C_{20}$ | 318.6 | 321.1 | 208.2 | 447.6 | 453.8 | -22.3 |
| $C_{19}N$ | 315.3 | 317.8 | 206.4 | 445.3 | 451.5 | -22.4 |
| $C_{18}N_2$-(1) | 320.8 | 323.3 | 215.9 | 448.6 | 454.9 | -5.4 |
| $C_{18}N_2$-(2) | 319.9 | 322.3 | 213.0 | 448.5 | 454.7 | -11.8 |
| $C_{18}N_2$-(3) | 320.0 | 322.4 | 214.8 | 448.0 | 454.3 | -6.9 |
| $C_{17}N_3$ | 316.0 | 318.5 | 208.2 | 445.6 | 451.9 | -18.4 |

Kinetic stability of the fullerene cages is associated with the possible defects that can appear at non-zero temperatures. According to the previous *ab initio* studies the Stone-Wales defect has the lowest activation energy barrier among the all defects considered in $C_{20}$ [16]. To analyze the formation of this defect we found the MEP connecting dodecahedral $C_{20}$ cage and its defective isomer (see Figure 2a). We obtained a single transition state on the MEP, its atomic structure is presented at Figure 2b.

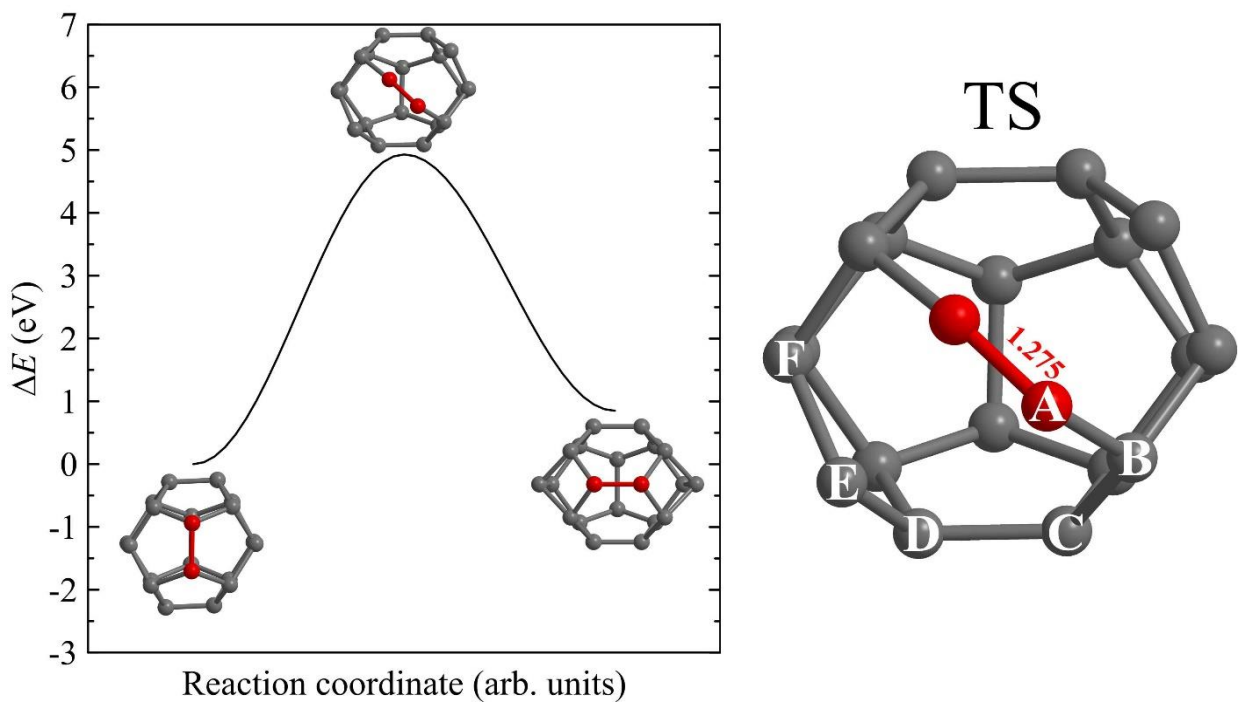

**Figure 2**. (left) Schematic MEP of the Stone-Wales defect formation in fullerene $C_{20}$ obtained by the NEB method. Rotating of C–C bond is highlighted in red. (right) Molecular structure of the transition state (saddle point).

The activation barrier $U$ for the Stone-Wales defect formation is assumed as the energy difference between the transition state configuration and the initial cage. According to our calculations $U$ is equal to 4.93 eV that is slightly higher than the value of 4.40 eV obtained earlier in Ref. [16] within the Hartree-Fock approach. We consider that the minor mismatch is associated with the various *ab initio* methods applied.

We suggest that the similar Stone-Wales defect can also appear in the nitrogen-doped cages through the same mechanism. To study the effect of substituting doping on the $U$ value we replaced a single carbon atom by a nitrogen one and after that, the nearest transition states were obtained. We probe six different sites for the substitution near the Stone-Wales defect labeled as A – F at Figure 2b. The energy differences between the corresponding transition states obtained and the dodecahedral $C_{19}N$ cage are (in eV) 3.81, 5.52, 5.35, 4.90, 2.98, and 4.51 for A, B, C, D, E, and F positions of nitrogen atom, respectively. Thus, the Stone-Wales transformations passing through the C–N bond rupture (A and E nitrogen positions) are more energetically favorable than those that are associated with C–C bonds breaking (B, C, D, and F nitrogen positions). The lowest value of *U* is achieved for the E position of nitrogen atom. Therefore, the corresponding isomerization trajectory has the highest probability. Applying the intrinsic reaction coordinate approach, we analyze this trajectory from the initial dodecahedral $C_{19}N$ to its defective isomer. We find the intermediate state on the reaction path that represents the cage with the 8-membered ring on its surface (see Figure 3a).

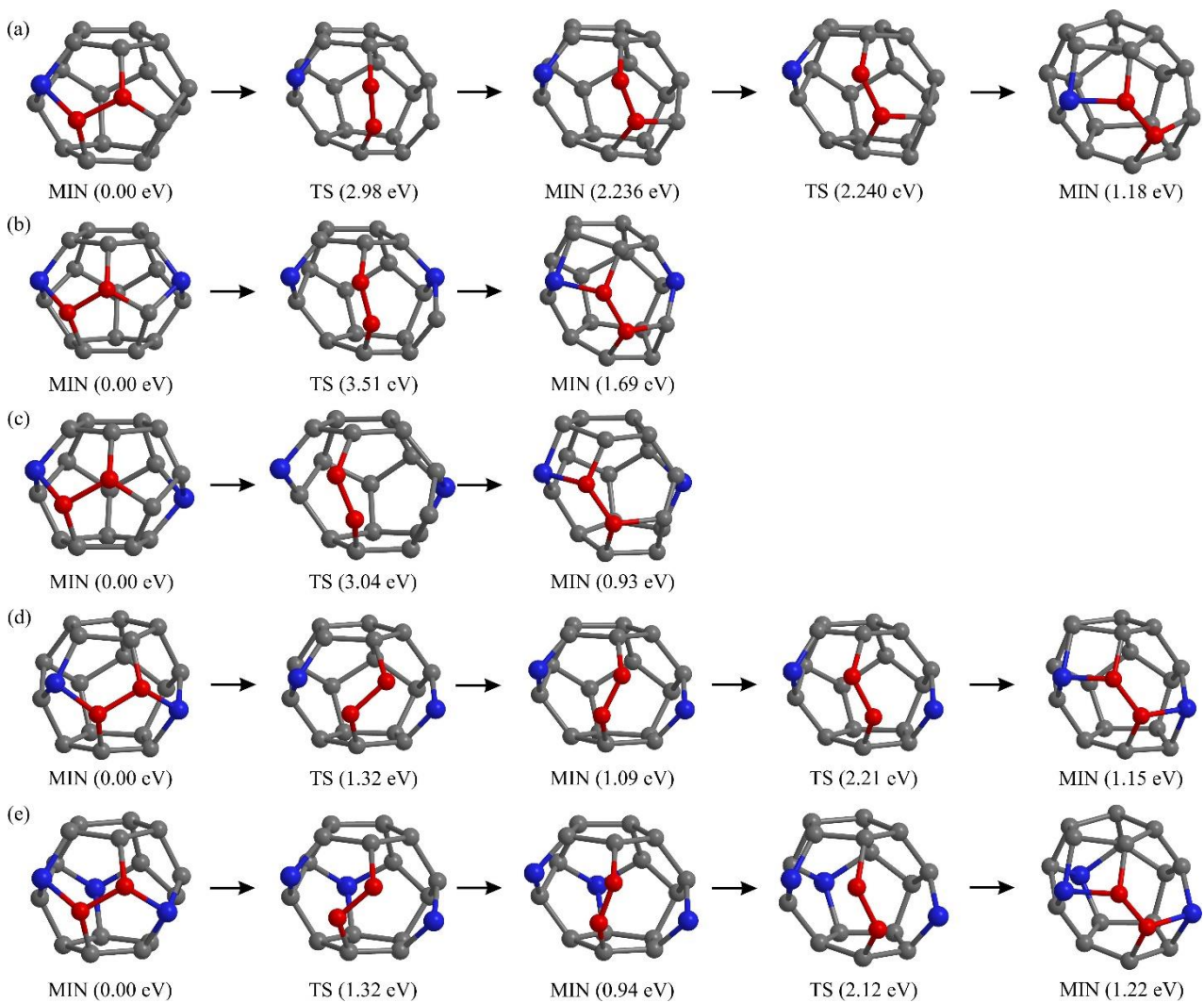

**Figure 3.** Mechanisms of the Stone-Wales defect formation in (a) $C_{19}N$, (b) $C_{18}N_2$-(1), (c) $C_{18}N_2$-(2), (d) $C_{18}N_2$-(3), and (e) $C_{17}N_3$ cages. Grey and blue colors denote the carbon and nitrogen atoms, respectively. Rotating of C–C bond is highlighted in red. Relative energies of local minima (MIN) and transition states (TS) are presented in brackets.

We fail to obtain the similar stable configuration for the non-substituted $C_{20}$ fullerene. Thus, only doped cages can form the stable isomer with the octagon ring on its surface (note that the similar partly open structures based on the nitrogen-doped $C_{60}$ fullerene were recently observed via the scanning tunneling microscopy [8]). On the other hand, the energy barrier separating such isomer from the more energetically favorable one with two squares on the surface is only 0.004 eV (see Figure 3a). This value is comparable with the accuracy of the density functional approach used. Therefore, the question of the intermediate state existence on the Stone-Wales isomerization path of $C_{19}N$ system remains open. In any case, this possible intermediate structure can hardly be observed experimentally.

Further, we regard cages with two doped nitrogen atoms. The Stone-Wales isomerization paths of $C_{18}N_2$-(1) and $C_{18}N_2$-(2) systems pass through the breaking of single C–N and C–C bonds, without any intermediate states (see Figures 3b and 3c). Opposite, in $C_{18}N_2$-(3) system two C–N bonds sequentially rupture, and the single well-defined intermediate configuration is observed (see Figure 3d).

Similar isomerization mechanism through the sequentially breaking of two C–N bonds is also observed for $C_{17}N_3$ system (see Figure 3e). The third additional nitrogen atom slightly reduces the isomerization barrier compared with the corresponding value for $C_{18}N_2$-(3) system. We consider that the similar isomerization paths take place for other $C_{17}N_3$ and more doped $C_{16}N_4$ isomers as well, excluding energetically unfavorable isomers containing adjacent or closely located nitrogens on their cages.

According to the Arrhenius law, the average time $\tau$ of the Stone-Wales defect formation at temperature $T$ can be evaluated as

$$\tau = (A \cdot g)^{-1} \exp(U/kT), \qquad (1)$$

where $k$ is the Boltzmann constant, $g$ is the degeneracy factor characterizing the number of the equivalent paths for defect formation (for example, $g = 30$ for the $C_{20}$ cage, since the defect can be introduced to any of 30 equivalent C–C bonds), and $A$ is the frequency factor that can be expressed as [33]

$$A = \frac{\prod_{i=1}^{3N-6} v_i}{\prod_{i=1}^{3N-7} v'_i}, \quad (2)$$

where $v_i$ are the eigenfrequencies of fullerene vibrations in the local minimum state, and $v'_i$ are the real frequencies of vibrations in the transition state (saddle point), $N = 20$ is the number of atoms in the molecular system considered. We estimate the times of Stone-Wales isomerization using the formulae (1) and (2) at room temperature ($T = 298$ K) and at $T = 750$ K. The latter is the typical temperature for annealing the nitrogen-doped fullerenes [8]. The times obtained as well as the data concerning the corresponding transition states are summarized in Table 2.

Table 2. Activation energy barriers $U$ (eV), imaginary eigenfrequencies of transition states $v_{TS}$ (s$^{-1}$), frequency factors $A$ (s$^{-1}$), degeneracy factors $g$ and overage estimated times $\tau$ (s) of the Stone-Wales isomerization in $C_{20-n}N_n$ ($n = 0 - 3$) fullerenes.

| System | $U$ | $v_{TS}$ | $A$ | $g$ | $\tau$ (298 K) | $\tau$ (750 K) |
|---|---|---|---|---|---|---|
| $C_{20}$ | 4.93 | 1.94·10$^{13}$ | 2.0·10$^{14}$ | 30 | 3.8·10$^{67}$ | 2.2·10$^{17}$ |
| $C_{19}N$ | 2.98 | 1.40·10$^{13}$ | 2.6·10$^{14}$ | 6 | 1.6·10$^{35}$ | 6.5·10$^{4}$ |
| $C_{18}N_2$-(1) | 3.51 | 0.96·10$^{13}$ | 3.2·10$^{15}$ | 4 | 1.8·10$^{43}$ | 3.0·10$^{7}$ |
| $C_{18}N_2$-(2) | 3.04 | 1.15·10$^{13}$ | 9.9·10$^{14}$ | 12 | 2.2·10$^{35}$ | 2.3·10$^{4}$ |
| $C_{18}N_2$-(3) | 2.21 | 1.40·10$^{13}$ | 1.3·10$^{15}$ | 1 | 1.8·10$^{22}$ | 5.4·10$^{-1}$ |
| $C_{17}N_3$ | 2.12 | 1.39·10$^{13}$ | 6.1·10$^{14}$ | 2 | 5.8·10$^{20}$ | 1.4·10$^{-1}$ |

As evident from Table 2, all nitrogen-doped dodecahedral cages considered retain their identities during the macroscopic times at room temperatures. However, defective isomers may appear at high temperatures typical for the fullerene annealing. Thus, we predict the presence of defective structures among the nitrogen-doped cages treated at $T \approx 750$ K. This fact should be taken into account for the interpreting of corresponding experimental data.

CONCLUSION

We presented the *ab initio* studies of Stone-Wales defect appearing mechanism in the pure and nitrogen-doped fullerenes $C_{20-n}N_n$ ($n = 0 - 3$). It was found that even a single nitrogen atom embedded into the fullerene cage instead of carbon atom induces the defect formation at high temperatures. The activation energy barrier of the Stone-Wales defect formation in $C_{19}N$ (2.98 eV) is sufficiently lower than the corresponding value in pure $C_{20}$ (4.93 eV). Further nitrogen doping leads to the subsequent reduction of the barrier. Depending on the number of nitrogen atoms and their relative position in the cage, Stone-Wales defect is formed through the additional intermediate state (local minimum) or directly. The C–N bonding in the nitrogen-doped system is found to be weaker than the C–C one. So, the Stone-Wales defect formation in $C_{20-n}N_n$ always associated with the breaking of one or two C–N bonds. All nitrogen-doped cages considered possess high kinetic stability at room temperature. However, in contrary to pure $C_{20}$ fullerene they become much less stable at $T \approx 750$ K that is typical value for the fullerene annealing. We suppose that the data obtained will be useful for the temperature regimes selection for the nitrogen-doped cages producing and synthesis.

ACKNOWLEDGEMENTS

The reported study was financially supported by Grant of the President of the Russian Federation, Grant No. MK-7410.2016.2.